\documentclass[twocolumn,twoside,slac_two]{revtex4}
\usepackage{graphicx}
\usepackage{fancyhdr}
\begin{document}

\title{APEX sub-mm monitoring of gamma-ray blazars\footnote{Based on
observations with the Atacama Pathfinder EXperiment (APEX) telescope in Chile.
The APEX telescope is operated by the Max-Planck-Institute f\"ur
Radioastronomie, the European Southern Observatory and the 
Onsala Space Observatory.}}

%

\author{S. Larsson}
\affiliation{Department of Astronomy and Department of Physics, 
Stockholm University, AlbaNova, SE-10691 Stockholm, Sweden, 
The Oskar Klein Centre for Cosmoparticle Physics, 
Stockholm, Sweden}
\author{L. Fuhrmann, A. Weiss, E. Angelakis, T. P. Krichbaum, I. Nestoras and J. A. Zensus on behalf of the F-GAMMA Collaboration}
\affiliation{Max-Planck-Institut f\"ur Radioastronomie, Auf dem H\"ugel 69, 53121 Bonn, Germany}
\author{M. Axelsson, D. Nilsson and F. Ryde}
\affiliation{Royal Institute of Technology, AlbaNova, SE-10691 Stockholm, Sweden, 
The Oskar Klein Centre for Cosmoparticle Physics, Stockholm, Sweden}
\author{L. Hjalmarsdotter}
\affiliation{Sternberg Astronomical Institute, 119992, Moscow, Russia}
\author{J. Larsson}
\affiliation{Department of Astronomy, 
Stockholm University, AlbaNova, SE-10691 Stockholm, Sweden, 
The Oskar Klein Centre for Cosmoparticle Physics, Stockholm, Sweden}
\author{A. Lundgren, F. Mac-Auliffe, R. Parra and G. Siringo}
\affiliation{European Southern Observatory, Vitacura Casilla 19001,
Santiago de Chile 19, Chile}

\begin{abstract}
  So far, no systematic long-term blazar monitoring programs
  and detailed variability studies exist at sub-mm wavelengths.
  Here, we present a new sub-mm blazar monitoring program 
  using the APEX 12-m telescope. A sample of about 40 $\gamma$-ray 
  blazars has been monitored since 2007/2008 with the LABOCA bolometer camera at 
  345\,GHz. First light curves, preliminary variability results and a first 
  comparison with the longer cm/mm bands (F-GAMMA program) are presented,
  demonstrating the extreme variability characteristics of blazars
  at such short wavelengths.

\end{abstract}

\maketitle

\thispagestyle{fancy}


\section{INTRODUCTION}

``Blazars'' comprise a sub-class of radio-loud AGN showing a
broad-band, double-humped spectral energy distribution (SED). They
differ from all other types of AGN chiefly because of their extreme
phenomelogical characteristics such as the extreme and broad-band flux
density and polarisation variability, high degree of polarisation,
fast superluminal motion and almost uniquely broad-band emission
characteristics including a bright and highly variable component of 
$\gamma$-ray emission~\cite{urry1999}. 

Despite many efforts over the last decades, several key questions
still remain unanswered in fully understanding the blazar phenomenon.
For instance: 
\begin{itemize}
\item which are the dominant, broad-band emission
processes involved (synchrotron self-Compton/inverse-Compton)? 
\item which mechanisms drive their often violent, broad-band
variability? (e.g. relativistic shocks, colliding plasma shells or
changing geometry due to helical/precessing jets; see e.g.~\cite{marscher1996}
 \cite{guetta1996} \cite{camenzind1992})
\item  what is the typical duty cycle of their activity? 
\item does the $\gamma$-ray emission originate from the 
base/foot-point of the jet or further out in the same shocked 
regions as the radio cm/mm/sub-mm band emission? 
\end{itemize}
Observations that can ascertain whether a relationship exists 
between the $\gamma$-ray and radio emission will be important
for the effort to answer these questions.

Consequently, variability studies furnish important clues about the
size, structure, physics and dynamics of the emitting region making
AGN/blazar monitoring programs of uttermost importance in providing
the necessary constraints for understanding the origin of energy
production. Here, observations at short-mm/sub-mm bands are crucial,
as they probe the innermost nuclear region and provide the important
direct link between the longer wavelength radio bands and the more 
energetic IR/optical to $\gamma$-ray regimes. 

Until now, there has been no systematic long-term blazar monitoring 
program and detailed variability study at sub-mm bands. 
Consequently, little is known about the variability
characteristics of AGN at sub-mm bands. Important parameters which
need to be determined are e.g. the variability
amplitude and time scales and how the variability is related to other 
bands, e.g. correlation strengths and time lags. The knowledge of the
variability behavior at sub-mm bands will thus better constrain the
modeling of the variability and spectral evolution in the synchrotron
branch of blazar SEDs.

In this framework, we initiated a blazar monitoring program using the
APEX sub-mm telescope. The aim of this effort, which is part of the 
F-GAMMA program~\cite{fuhrmann2007}~\cite{angelakis2008}, is to
perform the first long-term systematic study of blazar variability characteristics 
at sub-mm wavelengths. Here we present first light curves and some 
preliminary results on the relative variability for our sample of sources.
Details including the full first years of data and analysis will be 
presented in Fuhrmann et al. (in prep.).

\section{APEX observations with LABOCA}

The APEX 12\,m sub-mm telescope is located at 5100\,m altitude on
Llano Chajnantor, Chile. The observations are obtained with
the LABOCA camera~\cite{siringo2009}, which consists of 295 channels 
arranged in 9 concentric hexagons. LABOCA has a total field of view of
11.4 arcmin and allows observations in the 870 micrometer (345 GHz)
atmospheric window (bandwidth: 60 GHz).

\begin{figure*}
\centering
\vspace{-23mm}
\includegraphics[width=190mm]{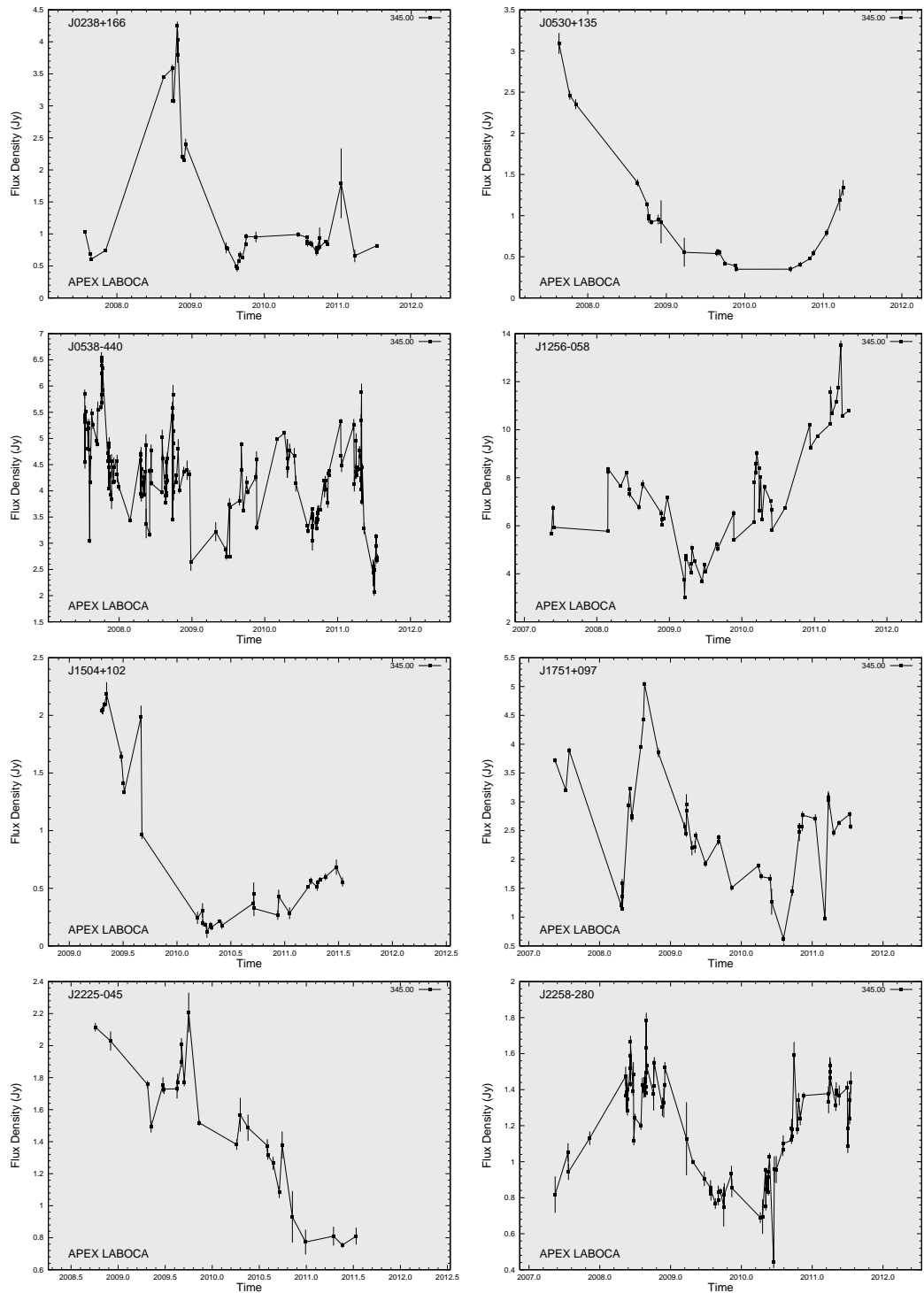}
\vspace{-20mm}
\caption{Examples of long-term light curves for several blazars monitored 
at sub-mm wavelengths with LABOCA on the APEX telescope. All sources
show strong variability in this band.} \label{mkfig}
\end{figure*}

The quasi-regular observations of our program started in 2008 and are
performed during several dedicated MPI, Swedish and ESO APEX LABOCA
time-blocks per year. In addition, we are aiming at denser, more
regular sampling through the regular and frequent pointing
observations performed at APEX since 2007 using our sources in the
framework of other projects and APEX technical time. Observations
within the F-GAMMA program are typically performed in {\it spiral
  observing mode} with a raster of four spirals of 20 or 35
seconds of integration each, depending on the source brightness at 345
GHz. At each run, skydip measurements for opacity correction and
frequent calibrator measurements are performed. The number of 
observations per source range from 2 to 177.

\begin{table}
\caption{List of monitored sources.
Source class
and 2FGL associations are taken from the second
LAT AGN Catalog~\cite{ackermann2011}.}
\begin{tabular}{|l|l|l|l|}
\hline
2FGL Name   &    Source Name       & Source Class  \\
\hline
J0210.7-5102   &    PKS 0208-512   & Blazar   \\   
J0217.9+0143   &    PKS 0215+015   & FSRQ   \\    
J0237.8+2846   &    4C +28.07      & FSRQ   \\   
J0238.7+1637   &    PKS 0235+164   & FSRQ   \\    
J0303.5-6209   &    PKS 0302-623   & FSRQ   \\    
J0339.4-0144   &    PKS 0336-01    & FSRQ    \\ 
J0403.9-3604   &    PKS 0402-362   & FSRQ   \\    
J0423.2-0120   &    PKS 0420-01    & FSRQ   \\    
J0530.8+1333   &    PKS 0528+134   & FSRQ   \\  
J0538.8-4405   &    PKS 0537-441   & BLLac   \\   
J0854.8+2005   &    OJ 287         & BLLac   \\  
J0909.1+0121   &    J0909+0121     & FSRQ   \\   
J1058.4+0133   &    4C +01.28      & BLLac   \\    
J1057.0-8004   &    PKS 1057-79    & BLLac   \\    
J1159.5+2914   &    J1159+292      & FSRQ   \\    
J1221.4+2814   &    W Comae        & BLLac   \\    
J1229.1+0202   &    3C 273         & FSRQ   \\    
J1256.1-0547   &    3C 279         & FSRQ   \\    
J1315.9-3339   &    PKS 1313-333   & FSRQ   \\    
J1325.6-4300   &    Cen A          & RG     \\  
J1428.0-4206   &    PKS 1424-418   & FSRQ   \\   
J1457.4-3540   &    PKS 1454-354   & FSRQ    \\ 
J1504.3+1029   &    PKS 1502+106   & FSRQ   \\    
J1512.8-0906   &    PKS 1510-08    & FSRQ   \\    
J1626.1-2948   &    PKS 1622-29    & FSRQ   \\   
J1635.2+3810   &    4C +38.41      & FSRQ   \\  
J1642.9+3949   &    3C 345         & FSRQ   \\  
J1653.9+3945   &    Mkn 501        & BLLac  \\   
J1733.1-1307   &    PKS 1730-13    & FSRQ   \\    
J1751.5+0938   &    PKS 1749+096   & BLLac   \\    
J1958.2-3848   &    PKS 1954-388   & FSRQ   \\    
J2056.2-4715   &    PKS 2052-47    & FSRQ   \\   
J2157.9-1501   &    PKS 2155-152   & FSRQ   \\    
J2158.8-3013   &    PKS 2155-304   & BLLac \\   
J2202.8+4216   &    BL Lac         & BLLac   \\    
J2225.6-0454   &    3C 446         & FSRQ   \\    
J2232.4+1143   &    CTA 102        & FSRQ   \\    
J2253.9+1609   &    3C 454.3       & FSRQ   \\    
J2258.0-2759   &    PKS 2255-282   & FSRQ    \\
\hline
\end{tabular}
\label{tab}
\end{table}

\section{The Sample} 
At APEX, a sub-sample of 25 prominent, famous, frequently active and
usually strong $\gamma$-ray blazars from the F-GAMMA sample is
observed together with a sample of 14 interesting southern
hemisphere $\gamma$-ray AGN. The complete list of APEX monitored
sources is given in Table 1.

\section{FIRST RESULTS: Sub-mm variability properties of gamma-ray AGN}
Some examples of source light curves are shown in Figure 1.
All the monitored sources, except one, show excess variability (over
that expected from measurement noise). The exception is Mkn 501,
which we exclude from the following variability analysis. For
all the remaining 38 sources the $\chi^2$-analysis gives a
probability of less than 0.1\% that the observed variability is
due to measurement noise. In a preliminary variability analysis, without 
taking the time sampling into 
account, the modulation index was calculated as $m = 100~\times~rms/mean$. 
For most sources the modulation index is in the range 10 - 50 $\%$, but there 
is a tail in the distribution which extends up to $m = 90 \%$ as shown
in Figure 2. The mean and median modulation indexes are 37 and 31\%
respectively. 
A number of sources show variations by a factor of 10 
or more between minimum and maximum flux. This is significantly 
larger than at the longer cm- and also short-mm bands. The variability 
is in general also faster and more directly correlated with the high 
energy emission\cite{larsson2012}. This is most likely an effect 
of opacity/synchrotron 
self-absorption increasing towards the longer radio wavelengths and 
indicates that the sub-mm emission regions are more co-spatial with 
the optical/$\gamma$-ray ones. 

In an analysis of the spectral evolution observed by the cm/mm F-GAMMA
program~\cite{angelakis2011}, it has been shown that the radio
flares and multi-frequency variability can be well described by only
two physical processes: (i) achromatic variability (possibly related
to helical or precessing jets) or (ii) evolving synchrotron flares or
shocks inside the relativistic jets (as described in 
e.g.~\cite{marscher1985}. 
In such synchrotron scenario of evolving AGN outbursts,
the flux-density variability first appears at higher frequencies
(IR/optical/UV/X-ray) and then propagates through the spectrum towards
longer wavelengths. The formation and evolution of shocks is expected 
to start at high synchrotron frequencies during their growth stage 
where the synchrotron self-absorption peak moves to lower frequencies 
while the peak flux and variability amplitude is increasing 
(mm/sub-mm bands) up to the plateau stage, followed by a subsequent 
decay stage (cm-bands). Here, the variability in the sub-mm band is
expected to be much more pronounced and faster than at longer cm-radio
bands\cite{valtaoja1992}. This was confirmed to 1\,mm wavelength 
by the earlier
observations (Fuhrmann et al. in prep.): the mean strength of variability 
(modulation index) steadily increases from 9.5\% at 110\,mm to 30\% at 
1\,mm. Our new LABOCA observations imply that this trend continues 
into the sub-mm band.  According to the three different regimes of 
shock evolution (growth, plateau, decay), such continued increase 
(or flattening) would then indicate that, on average, flares reach  
their plateau/decay phase at sub-mm bands well before they
do at cm bands. A more extensive analysis and discussion of the 
variability properties will be presented in a forthcoming paper.

\bigskip 
\begin{acknowledgments}
We are greatfull to the APEX staff for excellent support during
the many observations of this program. 

\end{acknowledgments}


\vspace*{2mm}

\begin{figure}[!b]
\centering
\includegraphics[width=80mm]{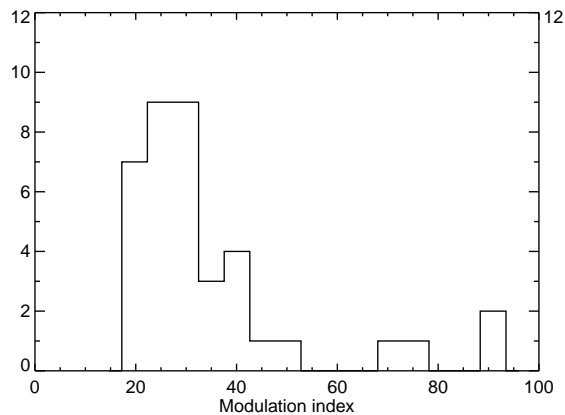}
\caption{The distribution of modulation index for the 38 
variable sources (excluding Mkn 501 for which the variability
is not significant). The bins in modulation index are 6 units wide.} 
\label{modindex}
\end{figure}

\end{document}